\documentclass[11pt]{article} 

\usepackage[utf8]{inputenc} 

\usepackage{geometry} 
\usepackage{graphicx} 

 \usepackage[parfill]{parskip} 

\usepackage{booktabs} 
\usepackage{array} 
\usepackage{paralist} 
\usepackage{verbatim} 
\usepackage{subfig} 

\usepackage{fancyhdr} 
\usepackage{sectsty}
\allsectionsfont{\sffamily\mdseries\upshape} 

\usepackage{amsmath, amssymb}

\usepackage{authblk}

\usepackage[style=alphabetic]{biblatex}
\usepackage{hyperref}

\title{A Note on the NP-Hardness of PARTITION Via First-Order Projections}
\author{Pa\'ul Risco Iturralde}
\affil{Independent researcher}
\date{} 

\begin{document}
\maketitle

\begin{abstract}
In the article ``On the (Non) NP-Hardness of Computing Circuit Complexity'', Murray and Williams imply the PARTITION  decision problem is not known to  be  NP-hard via $2^{n^{o(1)}}$-size AC0 reductions. In this note, we show PARTITION is NP-hard via first-order projections. Basically, we slightly modify well-known reductions from 3SAT to SUBSET-SUM and from SUBSET-SUM to PARTITION, but do so in the context of descriptive computational complexity, i.e., we use first-order logical formulas to define them. Hardness under  polynomial-size AC0 reductions follows because first-order reductions are a particular type of them. Thus, this note fills a gap in the literature.
\end{abstract}

\section{INTRODUCTION}

In their paper \cite{v013a004}, Murray and Williams mention ``$\text{TIME}\left(n^{o(1)}\right))$ reductions are powerful enough for almost all NP-competeness results, which have ``local'' structure transforming small pieces of the input to small pieces of the output,'' with the typical reduction from SUBSET-SUM to PARTITION being a potential counterexample (definition 1.1 in their paper explains what $\text{TIME}\left(t(n)\right))$ reductions are).  In footnote 2, they mention the straightforward reduction from SUBSET-SUM to PARTITION does not seem to be computable in $2^{n^{o(1)}}$-size AC0, because two numbers in the output of this reduction require taking the sum of all numbers in the input instance.

In this note, we prove PARTITION is NP-hard via  first-order projections (fops), which are reductions defined by restricted first-order logic formulas. This implies PARTITION is NP-hard under polynomial-size AC0 reductions \cite{doi:10.1137/S0097539794270236} \cite{doi:10.1137/0213028}.

How do we overcome the obstacle mentioned by Murray and Williams? First, let's recall the definitions of SUBSET-SUM, PARTITION, and the reduction itself \cite{10.5555/578533}.

\textbf{SUBSET-SUM:}

\textbf{Instance:} Finite set $B$, size $s(a) \in \mathbb{Z}^{+}$ for each $a \in B$, and positive integer $t$.

\textbf{Question:} Is there a subset $B' \subseteq B$ such that the sum of the sizes of the elements of $B'$ is exactly $t$?

\textbf{PARTITION:}

\textbf{Instance:} A finite set $A$ and a size $s(a) \in \mathbb{Z}^{+}$ for each $a \in A$.

\textbf{Question:} Is there a subset $A'\subseteq A$ such that
\[
\sum_{a\in A'}s(a) = \sum_{a\in A\setminus A'}s(a)? 
\]

The reduction takes an instance of SUBSET-SUM and outputs the set $A = B \cup \left\{b_{1}, b_{2}\right\}$, where
\begin{align*}
s(b_{1}) &= 2\left[\sum_{a \in A} s(a)\right] - t \text{ and}\\
s(b_{2}) &= \left[\sum_{a \in A} s(a)\right] + t.
\end{align*}

Now, some observations reveal that $s(b_{1})$ and $s(b_{2})$, while certainly dependent on the size of the input, can actually be made to not depend on the sizes of the members of $B$ as follows (the author first observed that PARTITION is NP-hard via polynomial-size AC0 reductions in \cite{170906}).

In \cite{10.1007/BF01195200}, Stewart shows that 3SAT, in which every instance has three distinct literals per clause, is NP-complete via fops. Sipser, in his book \cite{SipserToC3rd}, presents a (standard) reduction from 3SAT to SUBSET-SUM. That reduction is basically a projection, but assumes that a literal may occur more than once in a clause. Nevertheless, the same reduction works if we assume that every clause has three distinct literals.

Now, using the 3SAT vocabulary in Stewart's paper, every instance $\phi$ of 3SAT has $n$ clauses $c_{0}, c_{1}, \ldots , c_{n-1}$ and $n$ variables $x_{0}, x_{1}, \ldots , x_{n-1}$. Sipser's reduction creates a SUBSET-SUM instance in which $B$ is comprised of
\[
y_{0}, z_{0}, y_{1}, z_{1}, \ldots, y_{n-1}, z_{n-1} \quad \text{and} \quad g_{0}, h_{0}, g_{1}, h_{1}, \ldots, g_{n-1}, h_{n-1}.
\]
The decimal representation of the corresponding sizes, and target $t$, is like the one given in table \ref{table1}.

\begin{table}
\centering
\begin{tabular}{c | c c c c c c | c c c c c c} 
& 0 & 1 & 2 & 3 & $\ldots$ & $n-1$ & $c_{0}$ & $c_{1}$ & $c_{2}$  & $\ldots$ & $c_{n-1}$ \\
\hline
$y_{0}$ & 1 & 0 & 0 & 0 & $\ldots$ & 0 & 1 & 0 & 1  & $\ldots$ & 1 \\
$z_{0}$ & 1 & 0 & 0 & 0 & $\ldots$ & 0 & 0 & 1 & 1  & $\ldots$ & 0 \\
$y_{1}$ &  & 1 & 0 & 0 & $\ldots$ & 0 & 1 & 0 & 0  & $\ldots$ & 1 \\
$z_{1}$ &  & 1 & 0 & 0 & $\ldots$ & 0 & 1 & 1 & 0  & $\ldots$ & 0 \\
$y_{2}$ &  &  & 1 & 0 & $\ldots$ & 0 & 0 & 1 & 0  & $\ldots$ & 1 \\
$z_{2}$ &  &  & 1 & 0 & $\ldots$ & 0 & 0 & 0 & 1  & $\ldots$ & 0 \\
\vdots &  &  &  &  & $\ddots$ & $\vdots$ & $\vdots$ & &  & $\vdots$  & $\vdots$ \\
$y_{n-1}$ &  &  &  &  &  & 1 & 0 & 0 & 0  & $\ldots$ & 0 \\
$z_{n-1}$ &  &  &  &  &  & 1 & 0 & 0 & 0  & $\ldots$ & 0 \\
\hline
$g_{0}$ &  &  &  &  &  &  & 1 & 0 & 0  & $\ldots$ & 0 \\
$h_{0}$ &  &  &  &  &  &  & 1 & 0 & 0  & $\ldots$ & 0 \\
$g_{1}$ &  &  &  &  &  &  &  & 1 & 0  & $\ldots$ & 0 \\
$h_{1}$ &  &  &  &  &  &  &  & 1 & 0  & $\ldots$ & 0 \\
$\vdots$ &  &  &  &  &  &  &  &  &   & $\ddots$ & $\vdots$ \\
$g_{n-1}$ &  &  &  &  &  &  &  &  &   &  & 1 \\
$h_{n-1}$ &  &  &  &  &  &  &  &  &  &  & 1 \\
\hline
$t$ & 1 & 1 & 1 & 1 & $\ldots$ & 1 & 3 & 3 & 3  & $\ldots$ & 3
\end{tabular}
\caption{SUBSET-SUM instances in Sipser's reduction}
\label{table1}
\end{table}

For each $j \in \{0,1,\ldots,n-1\}$, there are 1s in the entries corresponding to the row-column pairs $(y_{j}, j)$, $(z_{j}, j)$, $(g_{j}, c_{j})$, and $(h_{j}, c_{j})$. Moreover, for $i,j \in \{0,1,\ldots,n-1\}$, there is a 1 in the $(y_{j},c_{i})$ entry if literal $x_{j}$ occurs in $c_{i}$,  and there is a 1 in the $(z_{j},c_{i})$ entry if literal $\neg x_{j}$ occurs in $c_{i}$. The rest of entries are 0s. Note also that target $t$ does not depend on the sizes of the members of $B$.

Now, observe that
\[
\sum_{a \in B} s(a) = 2222\ldots2555\ldots5,
\]
because each clause in $\phi$ has three distinct literals per clause. As a consequence,
\begin{align*}
2\left[\sum_{a \in B} s(a)\right] - t &= 3333\ldots3777\ldots7 \text{ and}\\
\left[\sum_{a \in B} s(a)\right] + t &= 3333\ldots3888\ldots8,
\end{align*}
i.e., $s(b_{1})$ and $s(b_{2})$ would not depend on the sizes of the members of $B$ if we assumed that every SUBSET-SUM instance had the form given in the table, but we can make such an assumption if we regard the problem as a restriction.

In the next section, we present the notions required to understand the document, and in the last one, we present our results.

\section{PRELIMINARIES}

We assume the reader is familiar with the basic concepts of mathematical logic as well as those of computational complexity theory. Also, we assume acquaintance with elementary circuit complexity.

In what follows, we present the basics of descriptive complexity. The reader requiring more details may consult \cite{ImmermanDC}. 

In descriptive complexity, tools from mathematical logic are used to study concepts and establish results in computational complexity theory. As an example, any problem in NP may be described by a second-order exisential sentence, that is, a logical sentence $\Phi \in \text{SO}\exists$. Consider, for instance, the SAT decision problem. In order to define this problem with a logical sentence, we first need a vocabulary. Let $<P^{2},N^{2}>$ be a tuple consisting of the binary relation symbols $P^{2}$ and $N^{2}$. We interpret such a vocabulary through structures $\mathcal{A} = <|\mathcal{A}|, P^{\mathcal{A}}, N^{\mathcal{A}}>$, where $|\mathcal{A}|$, called the \emph{universe of structure} $\mathcal{A}$, is an initial segment of the natural numbers, $\{0,1,\ldots, n-1\}$, where $n \geq 2$, and $P^{\mathcal{A}}$ and $N^{\mathcal{A}}$ are binary relations over that set, such that
\begin{align*}
(j,i) \in P^{\mathcal{A}} \quad \text{if and only if} & \quad \text{literal} \, x_{j} \, \text{occurs in clause} \, c_{i}, \, \text{and}\\
(j,i) \in N^{\mathcal{A}} \quad \text{if and only if}& \quad \text{literal} \, \neg x_{j} \, \text{occurs in clause} \, c_{i}.
\end{align*}
Here, the $x_{j}$ and $c_{i}$ are, respectively, the variables and clauses in $\varphi$, the SAT instance.

We will refer to the cardinality of $|\mathcal{A}| = \{0,1,\ldots, n-1\}$ as the \emph{size of structure} $\mathcal{A}$.

From the previous paragraph, a structure $\mathcal{A}$ may be thought of as an instance of the SAT problem, $\varphi$, with $n$ variables and $n$ clauses. So, if we wanted to represent a formula such as
\[
\varphi \quad \equiv \quad (\neg x_{1} \lor x_{0} \lor x_{2}) \, \land \, (\neg x_{2} \lor x_{0})
\]
in this setting, we would pick $n = 3$ and would declare
\begin{align*}
P^{\mathcal{A}}&=\left\{(0,0), (2,0), (0,1),(2,2)\right\} \quad \text{and}\\
N^{\mathcal{A}}&=\left\{1,0), (2,1),(2,2)\right\}.
\end{align*}
Note that $(2,2) \in N^{\mathcal{A}}, P^{\mathcal{A}}$, because we don't want to have empty clauses.

We abbreviate expressions such as $(j,i) \in P^{\mathcal{A}}$ with $P(j,i)$.

Now, with this vocabulary, a formula defining SAT is
\[
\Phi_{SAT} \quad \equiv \quad (\exists S)(\forall x)(\exists y)((y \in S \, \land\, P(y,x)) \, \lor \, (y \notin S \, \land\, N(y,x)));
\]
that is, $\mathcal{A} \models \Phi_{SAT}$ if and only if $\mathcal{A}$ is a positive instance of SAT. The symbol $S$ captures the notion of a truth assignment to the variables of the formula: $x_{j}$ is assigned $1$ if and only if $j$ is in the set that $S$ represents. Note that $\Phi_{SAT} \in \text{SO}\exists$ because it is a second-order sentence in which the relation variable is quantified by an existential quantifier.

If a binary input relation symbol, say $W^{2}$, is present in a vocabulary, we can think of $W^{\mathcal{A}}$ as a binary matrix whose $(j,i)$-entry is $1$ if and only if $(j,i) \in W^{\mathcal{A}}$. When we use terms such as  ``the $W$ matrix'', we're referring to this matrix.

In descriptive complexity, we refer to decision problems as \emph{boolean queries}. More specifically, a boolean query is the set of structures corresponding to the positive instances of a problem. The set of all structures with a given vocabuary $\tau$ is denoted by \\ $\text{STRUC}[\tau]$.

Even though the vocabularies we'll consider in this note don't exhibit them, the following symbols are also part of these: $=, \leq, \text{SUC, PLUS, TIMES}, 0, 1, max$. The last three are referred to as \emph{numeric constant symbols}, while the others are called \emph{numeric relation symbols.} On the other hand, the \emph{input relation symbols} and \emph{input constant symbols} are the ones not in the previous list so, for instance, $P$ and $N$ above are input relation symbols.

\emph{Numeric formulas} are formulas in which no input relation symbol occurs.

Now, we provide a couple of definitions:
\begin{align*}
\text{SUC}(i,j) \quad \text{if and only if} & \quad j \, \, \text{is the succesor of} \, \, i \, \, \text{and}\\
\text{PLUS}(i,j,k) \quad \text{if and only if} & \quad i+j=k.
\end{align*}
The numeric constant symbols 0, 1, and $max$ will be interpreted, respectively, as the numbers 0, 1, and $n-1$. Finally, $=$ and $\leq$ will be given their usual meanings.

Now, the standard notion of reduction in the descriptive setting is that of a first-order reduction, that is, a many-one reduction which can be defined by a first-order logical formula.

Let $\sigma$ and
\[
\tau = <R_{1}^{a_{1}}, R_{2}^{a_{2}}, \ldots, R_{r}^{a_{r}}, c_{1}, c_{2}, \ldots ,c_{s}>
\]
be two vocabularies, where the $R_{i}^{a_{i}}$ are input relation symbols of arity $a_{i}$, and the $c_{j}$ are input constant symbols. If $S$ and $T$ are boolean queries whose underlying vocabularies are, respectively, $\sigma$ and $\tau$, a $k$-\emph{ary first-order reduction from} $S$ to $T$ is a function $I: \text{STRUC}[\sigma] \to \text{STRUC}[\tau]$ defined by a tuple of first-order formulas with vocabulary $\sigma$,
\[
<\varphi_{0}, \varphi_{1}, \varphi_{2}, \ldots, \varphi_{r}, \psi_{1}, \psi_{2}, \ldots, \psi_{s}>,
\]
such that for all $\mathcal{A} \in \text{STRUC}[\sigma]$,
\[
\mathcal{A} \in S \quad \text{if and only if} \quad I(\mathcal{A}) \in T,
\]
where
\[
I(\mathcal{A}) = <\left|I(\mathcal{A})\right|, R_{1}^{I(\mathcal{A})}, R_{2}^{I(\mathcal{A})}, \ldots, R_{r}^{I(\mathcal{A})}, c_{1}^{I(\mathcal{A})}, c_{2}^{I(\mathcal{A})}, \ldots, c_{s}^{I(\mathcal{A})}> \]
is given by:
\begin{equation*}
\left|I\left(\mathcal{A}\right)\right| = \left\{\left(b_{1}, \ldots, b_{k}\right) \in |\mathcal{A}|^{k}: \, \mathcal{A} \models \varphi_{0}\left(b_{1},\ldots,b_{k} \right)\right\},
\end{equation*}
\begin{equation*}
R_{i}^{I(\mathcal{A})} = \left\{\left(d_{1}, \ldots, d_{a_{i}}\right) \in  |I(\mathcal{A})|^{a_{i}}: \, \mathcal{A} \models \varphi_{i}\left(d_{1}, \ldots, d_{a_{i}}\right)\right\}
\end{equation*}
for $i= 1, \ldots, r$, and with $c_{j}^{I(\mathcal{A})}$ the unique $\left(b_{1}, \ldots, b_{k}\right) \in \left|I\left(\mathcal{A}\right)\right| $ such that\\ $\mathcal{A} \models \psi_{j}\left(b_{1},\ldots,b_{k} \right)$, this for $j= 1, \ldots, s$.

Now, a first-order projection is a first-order reduction in which the formulas $\varphi_{i}$, for $i=1,2,\ldots,r$, and the formulas $\psi_{j}$ have the form
\[
\alpha_{0} \, \lor \, (\alpha_{1}\land \lambda_{1}) \, \lor \, (\alpha_{2}\land \lambda_{2}) \ldots (\alpha_{s}\land \lambda_{s}),
\]
where the $\alpha_{e}$ are numeric, pairwise mutually exclusive formulas, and the $\lambda_{e}$ are literals whose underlying relation symbols are input relation symbols. Moreover, $\varphi_{0},$ the formula defining $|I(\mathcal{A})|$, is a numeric formula.

If there is a first-order projection reducing boolean query $S$ to boolean query $T,$ we write $S \leq_{fop} T$. The $\leq_{fop}$ relation is transitive.

As observed in \cite{doi:10.1137/S0097539794270236}, NP-hardness under first-order reductions implies nonmembership in the class AC0. Moreover, by  Razborov-Smolensky's theorem, NP-hardness in this sense guarantees that a problem does not have $\text{AC0}[p]$ circuits for any prime $p$ (otherwise, MAJORITY would have such circuits).

\section{RESULTS}

We present reductions showing 3SAT $\leq_{fop}$ SUBSET-SUM and SUBSET-SUM $\leq_{fop}$ PARTITION. The result will follow because 3SAT (in which every instance has three distinct literals per clause) is NP-hard via fops. Our reductions work because, basically, we're adding columns and rows of 0s between the ones already present in the standard reductions (refer to table \ref{table1}) and, therefore, the sums of the entries in the nonzero columns are preserved.

The 3SAT vocabulary used in Stewart's paper is $<P^{2},N^{2}>$. $P(j,i)$ if and only if variable $x_{j}$ occurs in clause $c_{i}$, and $N(j,i)$ if and only if literal  $\neg x_{j}$ occurs in $c_{i}$. Here, every clause in each instance is assumed to have three distinct literals.

On the other hand, the vocabulary for SUBSET-SUM we use is $<W^{2},L^{1}>$. $W(i,j)$ if and only if the $j$-th bit of $s(i)$ is 1, where the least siginificant bit is the one in the $(n-1)$-th position (assuming the universe of the structure has size $n$.) $L(j)$ means the $j$-th bit of target $t$ is 1.

With the way we interpret $W$, the $i$-th row of the binary matrix associated to it, is precisely the size of element $i$, $s(i)$, in binary.

The fop we will define, $\rho_{1}$, is 4-ary. First, we provide a sketch of our reduction, and then we present the corresponding formulas. If the universe of our input structure has size $n$, the universe of the corresponding output structure will have size $n^{4}$. To be more precise, the elements of the universe will be $4$-tuples in which each entry is an element of $\{0,1, \ldots, n-1\}$. The order attached to the output structure is the lexicographic order.

Now, if $j$ is one of the columns in the first half of table \ref{table1}, we'll assign it the tag $(0,j,1,n-1)$. Rows $y_{j}$ and $z_{j}$ will be tagged, respectively, with $(0,j,1,n-2)$ and $(0,j,1,n-1)$. Now, column $c_{i}$ will be identified with $(n-1,i,1,n-1)$, and rows $g_{i}$ and $h_{i}$ with $(n-1,i,1,n-2)$ and $(n-1,i,1,n-1)$, respectively. The fact that column $j$ and row $z_{j}$, as well as column $c_{i}$ and  row $h_{i}$, have the same tags will be key in our second reduction.

With the tags in place, it is easy to the define the 1s corresponding to the row-column pairs $(y_{j},j)$ and $(z_{j},j)$ as well as the 1s corresponding to the $(g_{i},c_{i})$ and $(h_{i},c_{i})$ pairs. The remaining 1s are defined by using $P(j,i)$ and $N(j,i)$ in addition to the tags.

It's not hard to verify that for all $n\geq 2$, there are at least three columns of 0s between the nonzero columns. Remember that we use the lexicographic order for $4$-tuples. So, for instance, between consecuitve nonzero columns $(n-1,i-1,1,n-1)$ and $(n-1,i,1,n-1)$, lie the ones with tags $(n-1,i,0,0), (n-1,i,0,1),$ and $(n-1,i,1,0)$, which are columns of 0s. This is important as we want to write the 3s in $t=1111\ldots1333\ldots3$ in binary; that is, when we write the target, we want to have 1s in columns $(n-1,i,1,n-2)$ (which is a column of 0s) and $(n-1,i,1,n-1)$.

Let's define $\rho_{1}$.

$\varphi_{0}\equiv \mathbf{true}$.

$\varphi_{1}\left(x_{1},x_{2},x_{3},x_{4},w_{1},w_{2},w_{3},w_{4}\right)$ defines $W^{\rho_{1}(\mathcal{A})}$ for each structure $\mathcal{A}$ with vocabulary $<P^{2},N^{2}>$.

1s in row-column pairs $(y_{j},j)$ (that is, the $((0,j,1,n-2),(0,j,1,n-1))$ row-column pairs):
\begin{align*}
\alpha_{00} \equiv & \quad  x_{1}=0 \, \land \, x_{3}=1 \, \land \, \text{SUC}\left(x_{4},max\right) \quad \land\\
& \quad w_{1}=0 \, \land \, w_{2}=x_{2} \, \land \, w_{3}=1 \, \land \, w_{4}=max
\end{align*}

1s in row-column pairs $(z_{j},j)$:
\begin{align*}
\alpha_{01} \equiv & \quad  x_{1}=0 \, \land \, x_{3}=1 \, \land \, x_{4}=max \quad \land\\
& \quad w_{1}=0 \, \land \, w_{2}=x_{2} \, \land \, w_{3}=1 \, \land \, w_{4}=max
\end{align*}

1s in row-column pairs $(g_{i},c_{i})$:
\begin{align*}
\alpha_{02} \equiv & \quad  x_{1}=max \, \land \, x_{3}=1 \, \land \, \text{SUC}\left(x_{4},max\right) \quad \land\\
& \quad w_{1}=max \, \land \, w_{2}=x_{2} \, \land \, w_{3}=1 \, \land \, w_{4}=max
\end{align*}

1s in row-column pairs $(h_{i},c_{i})$:
\begin{align*}
\alpha_{03} \equiv & \quad  x_{1}=max \, \land \, x_{3}=1 \, \land \, x_{4}=max \quad \land\\
& \quad w_{1}=max \, \land \, w_{2}=x_{2} \, \land \, w_{3}=1 \, \land \, w_{4}=max
\end{align*}

1s in row-column pairs $(y_{j},c_{i})$:
\begin{align*}
\alpha_{1}\land \lambda_{1} \equiv & \quad  x_{1}=0 \, \land \, x_{3}=1 \, \land \, \text{SUC}\left(x_{4},max\right) \quad \land\\
& \quad w_{1}=max \, \land \, w_{3}=1 \, \land \, w_{4}=max  \quad \land\\
& \quad P\left(x_{2},w_{2}\right)
\end{align*}

1s in row-column pairs $(z_{j},c_{i})$:
\begin{align*}
\alpha_{2}\land \lambda_{2} \equiv & \quad  x_{1}=0 \, \land \, x_{3}=1 \, \land \, x_{4}=max \quad \land\\
& \quad w_{1}=max \, \land \, w_{3}=1 \, \land \, w_{4}=max  \quad \land\\
& \quad N\left(x_{2},w_{2}\right)
\end{align*}

Letting
\[
\alpha_{0} \equiv \alpha_{00} \lor \alpha_{01} \lor \alpha_{02}  \lor \alpha_{03},
\]
we arrive at
\[
\varphi_{1} \equiv \alpha_{0} \, \lor \, (\alpha_{1}\land \lambda_{1}) \, \lor \, (\alpha_{2}\land \lambda_{2}).
\]
It's not hard to see this formula satisifies the properties of a projective formula.

$\varphi_{2}\left(w_{1},w_{2},w_{3},w_{4}\right)$ defines $L^{\rho_{1}(\mathcal{A})}$. Recall, target $t=11\ldots133\ldots3$, so in the case of the 1s, we just need to locate the corresponding columns, and in the case of the 3s, we need to write a 1 to the left of each 1 corresponding to the $c_{i}$-columns. In conclusion, there must be 1s in columns $(0,j,1,n-1)$, $(n-1,i,1,n-1)$ and $(n-1,i,1,n-2)$.
\begin{align*}
\varphi_{2} \equiv & \quad  w_{1}=0 \, \land \, w_{3}=1 \, \land \, w_{4}=max \quad \lor\\
& \quad  w_{1}=max \, \land \, w_{3}=1 \, \land \, w_{4}=max \quad \lor\\
& \quad  w_{1}=max \, \land \, w_{3}=1 \, \land \, \text{SUC}\left(w_{4},max\right)
\end{align*}
So $\varphi_{2}$, being a numeric formula, satisfies the desired properties.

Now, we explain why the reduction works. Note that target $t$ in our reduction isn't exactly the same as target $t$ in the original table \ref{table1}, because there are additional columns of 0s. We believe, though, readers will find it easier to follow the proof if they refer to the original table.

Suppose there exists a truth assignment to the 3SAT instance satisfying every clause in it. Let's construct the set $B'$ in the corresponding SUBSET-SUM instance. If variable $x_{j}$ is assigned 1, put $y_{j}$ in $B'$, that is, put the element corresponding to row $(0,j,1,n-2)$ in $B'$; otherwise, put $z_{j}$, the element corresponding to row $(0,j,1,n-1)$, in $B'$. If we add what we have so far, there will be a 1 in every bit position corresponding to columns $(0,j,1,n-1)$.

Now, given that every clause is satisfied, the current sum on any column with tag $(n-1,i,1,n-1)$ (corresponding to $c_{i}$) is an integer beween 1 and 3; if it's not 3, we add $g_{i}$ or $h_{i}$ to $B'$ so as to make it 3. Because 11 is 3 in binary, there are now 1s in columns $(n-1,i,1,n-2)$ and $(n-1,i,1,n-1)$ of our sum. This sum is precisely target $t$.

Conversely, assume there exists $B'\subseteq B$ such that the sum of the sizes of its elements equals target $t$. That means exaclty one of $y_{j}$ and $z_{j}$ is part of $B'$ for each $j$. If $y_{j} \in B'$, we assign $x_{j}$ the value 1; otherwise, we assign it 0. Let $c_{i}$ be one of the clauses of the 3cnf-formula. Note its corresponding column is the one with tag $(n-1,i,1,n-1)$. There are exactly five 1s in this column. On the other hand, target $t$ has 1s in columns $(n-1,i,1,n-2)$ and $(n-1,i,1,n-1)$. So, exactly three 1s in column $(n-1,i,1,n-1)$ make up this sum; at most two of them can come from sizes $g_{i}$ and $h_{i}$, so at least one comes from $y_{j}$ or $z_{j}$ for some $j$. If it comes from the former, $x_{j}$ is assigned 1 and  occurs in $c_{i}$; otherwise, $x_{j}$ is assigned 0 and $\neg x_{j}$ occurs in $c_{i}$. In both cases, $c_{i}$ is satisfied.

In conclusion, we have proved that a restricition of SUBSET-SUM is NP-hard via fops. In the next section, we'll assume every instance has the form in  this restriction.

\subsection{PARTITION IS NP-HARD VIA FOPS}

Now, we present the proof of SUBSET-SUM $\leq_{fop}$ PARTITION. The vocabulary used for PARTITION is $<T^{2}>$; then, $T(i,j)$ is interpreted in the same way as $W(i,j)$. Basically, we'll add columns and rows of 0s to the $W$ matrix in the SUBSET-SUM instances output by reduction $\rho_{1}$, besides creating sizes $s(b_{1})$ and $s(b_{2})$.

We'll assume the instances of SUBSET-SUM have the same form as the ones in the image of $\rho_{1}$; in particular, note that $k$ is a  nonzero column of the square binary matrix associated to $W$ if and only if $W(k,k)$; this is because of the tags assigned to the $(z_{j},j)$ and $(h_{i},c_{i})$ row-column pairs. Also note that every structure in the image of $\rho_{1}$ has size $n\geq 16$; the reason is $\rho_{1}$ is $4$-ary and every instance (structure) of 3SAT has size greater than or equal to $2$. As a consequence, the universe of any structure in the SUBSET-SUM restriction has $4,3,$ and $2$ as members.

Our fop, $\rho_{2}$, is 2-ary. We proceed using the standard reduction. We define $T^{\rho_{2}\left(\mathcal{B}\right)}$ for each structure $\mathcal{B}$ with vocabulary $<W^{2},L^{1}>$ in the SUBSET-SUM restriction, by cases. First, we copy the 1s of the $W$ matrix. In the next stages, we define $s(b_{1})$ and $s(b_{2})$.

Formulas defining $\rho_{2}$:

$\varphi_{0}\equiv \mathbf{true}$.

$\varphi_{1}(x_{1},x_{2},w_{1},w_{2})$ defines $T^{\rho_{2}(\mathcal{B})}$.

Copying the 1s of the $W$ matrix to the $T$ matrix:
\[
x_{2}=0 \, \land \, w_{2}=4 \, \land \, W(x_{1}, w_{1})
\]
So, if there was a 1 in row $i$ and column $j$ of the $W$ matrix, that 1 is copied to the entry  in row $(i,0)$ and column $(j,4)$ of the $T$ matrix.

Up to this point, there are several columns in the $T$ matrix which are nonzero columns. Moreover, between these there at least four columns of 0s. Indeed, if $k < j$ and $(j,4)$ and $(k,4)$ are two consecutive nonzero columns, then columns $(j,3), (j,2), (j,1),$ and $(j,0)$ contain only 0s and lie between these.

We now switch focus to $s(b_{1})$. The row in which this size will be placed is $(0,1)$. Note this row is different from the rows in the previou stage. Recall $s(b_{1}) = 3333\ldots3777\ldots7$ in the standard reduction. Let's first handle the $3$s. $3$ is a 1 followed by another 1 in binary.

If there is a $1$ in column $(j,4)$ of the $T$ matrix, that stems from the previous stage, we want to have $1$s in the $(0,1,j,4)$ and $(0,1,j,3)$ row-comlun pairs. The following formula captures our intentions. Here we use the fact that $k$ is a nonzero column of the $W$ matrix if and only if there is a $1$ in the $(k,k)$ entry of that matrix. Also, note the $3$s in $3333\ldots3777\ldots7$ happen to be in the left half of the decimal representation of $s(b_{1})$. 
\begin{align*}
x_{1}=0 \, \land \, x_{2}=1\, \land \, w_{2}=4 \, \land \, \beta_{left-half}(w_{1}) & \land \,W(w_{1},w_{1}) \quad \lor \\ 
x_{1}=0 \, \land \, x_{2}=1\, \land \, w_{2}=3 \, \land \, \beta_{left-half}(w_{1}) & \land \,W(w_{1},w_{1}) 
\end{align*}
We will eventually define $\beta_{left-half}(w_{1})$.

Now, let's handle the 7s in $s(b_{1})=3333\ldots3777\ldots7.$ 7 is 111 in binary, so we now want to put three 1s in consecutive columns. The following formula achives this, the reason being analogous to the one given above.
\begin{align*}
x_{1}=0 \, \land \, x_{2}=1\, \land \, w_{2}=4 \, \land \, \beta_{right-half}(w_{1}) & \land \,W(w_{1},w_{1}) \quad \lor \\
x_{1}=0 \, \land \, x_{2}=1\, \land \, w_{2}=3 \, \land \, \beta_{right-half}(w_{1}) & \land \,W(w_{1},w_{1}) \quad \lor \\
x_{1}=0 \, \land \, x_{2}=1\, \land \, w_{2}=2 \, \land \, \beta_{right-half}(w_{1}) & \land \,W(w_{1},w_{1})
\end{align*}

Using similar strategies, we now define $s(b_{2})$. The correspnding row is the one with tag $(1,1).$ In the original reduction,  $s(b_{2})=3333\ldots3888\ldots8.$ Here's how we define the 3s.
\begin{align*}
x_{1}=1 \, \land \, x_{2}=1\, \land \, w_{2}=4 \, \land \, \beta_{left-half}(w_{1}) & \land \,W(w_{1},w_{1}) \quad \lor \\ 
x_{1}=1 \, \land \, x_{2}=1\, \land \, w_{2}=3 \, \land \, \beta_{left-half}(w_{1}) & \land \,W(w_{1},w_{1})
\end{align*}

Now, let's define the 8s (recall 8 is 1000 in binary). 
\[
x_{1}=1\, \land \, x_{2}=1\, \land \, w_{2}=1 \, \land \, \beta_{right-half}(w_{1}) \, \land \,W(w_{1},w_{1})
\]

Note the formula $w_{2}=3$ above is of course an abbreviation of
\[
\left(\exists t_{1}\right)\left(\text{SUC}\left(1,t_{1}\right) \, \land \, \text{SUC}\left(t_{1},w_{2}\right)\right)
\]
or equivalent numeric first-order formulas. The same applies to related formulas.

Now, $\beta_{left-half}(k)$ is satisfied if and only if $k \leq \frac{n-2}{2}$ or $k < \frac{n-1}{2}$, where the fractions are integers. Here's the definition of the formula.
\begin{align*}
\beta_{left-half}(w_{1})\equiv& \left(\exists t_{1}\right)\left(\exists t_{2}\right)\left(\text{SUC}\left(t_{1},max\right)\, \land \, \text{PLUS}\left(t_{2},t_{2},t_{1}\right)\, \land \, w_{1} \leq t_{2}\right) \quad \lor\\
& \left(\exists t_{2}\right)\left(\text{PLUS}\left(t_{2},t_{2},max\right)\, \land \, w_{1}< t_{2}\right)
\end{align*}

$\beta_{right-half}(k)$ is satisfied if and only if $k > \frac{n-2}{2}$ or $k \geq \frac{n-1}{2}$. Here's the definition.
\begin{align*}
\beta_{right-half}(w_{1})\equiv& \left(\exists t_{1}\right)\left(\exists t_{2}\right)\left(\text{SUC}\left(t_{1},max\right)\, \land \, \text{PLUS}\left(t_{2},t_{2},t_{1}\right)\, \land \, w_{1} > t_{2}\right) \quad \lor\\
& \left(\exists t_{2}\right)\left(\text{PLUS}\left(t_{2},t_{2},max\right)\, \land \, w_{1} \geq t_{2}\right)
\end{align*}

Note that $\beta_{left-half}(w_{1})$ and $\beta_{right-half}(w_{1})$ are mutually exclusive formulas.

In conclusion, the formula defining the reduction can be expressed as
\[
\varphi_{1} \equiv \alpha_{0} \, \lor \, \alpha_{1}\land \lambda_{1} \, \lor \, \alpha_{2} \land \lambda_{2},
\]
where $\alpha_{0} \equiv \mathbf{false}$, $\lambda_{1}$ is $W(x_{1}, w_{1})$, and $\lambda_{2}$ is $W(w_{1},w_{1})$. $ \alpha_{2}$ is the disjunction of the numeric subformulas used in defining $s(b_{1})$ and $s(b_{2})$. It's not hard to see $\alpha_{1}$ and $\alpha_{2}$ are mutually exclusive.

The way we'll prove this reduction works is by computing the sums of the entries in the nonzero columns.

Before proceeding, define $F: E \subseteq B \to D\subseteq A$ as the function that maps the elements in the SUBSET-SUM instance with nonzero sizes to the corresponding elements of the PARTITION instance. The latter are the elements generated in the first stage of the reduction.

Now, let's assume there exists $B' \subseteq B$ in the SUBSET-SUM instance, such that the sum of the sizes of the elements in $B'$ is exactly $t$. We want to prove the instance output by our reduction is a positive instance of PARTITION. Define $(A',A \setminus A')$ as the partition of $A$ (in the PARTITION instance), with
\[
A' = F \left(E \cap B'\right) \cup \left\{b_{1}\right\}.
\]
Thus, the sum of the sizes of the elements of $A'$ is $t+s(b_{1})$:
\begin{center}
\resizebox{\textwidth}{!}{\begin{tabular}{ c c c c c c c c c c c c c c c c c c c c c c c c c c c }
& &  &1 & \ldots & &  & 1 & \ldots & &  & 1 & \ldots & &  & 1 & 1 &  \ldots & &  & 1 & 1 & \ldots & & & 1 &1 \\ 
+ & & & & & & & & & & & & & & & & & & & & & & & & & &  \\ 
&  & 1 & 1 & \ldots &  & 1 &1 & \ldots &  & 1 & 1 & \ldots &  & 1 & 1 & 1 & \ldots &  & 1 & 1 & 1 & \ldots &  & 1 & 1 & 1 \\
\hline
& 1& 0 & 0 & \ldots & 1 & 0 &0 & \ldots & 1 & 0 & 0 & \ldots & 1 & 0 & 1 & 0 & \ldots & 1 & 0 & 1 & 0 & \ldots & 1 & 0 & 1 & 0.
\end{tabular}}
\end{center}

Note that the sum of the sizes in the PARTITION instance is given by
\[
\sum_{a \in F(E)} s(a) \quad + s(b_{1}) \quad + s(b_{2}).
\]
In the original reduction, the left sum equals $2222\ldots2555\ldots5$, which stems from adding all the sizes in the SUBSET-SUM instance. So the sum above in the corresponding order is:
\begin{center}
\resizebox{\textwidth}{!}{\begin{tabular}{ c c c c c c c c c c c c c c c c c c c c c c c c c c c c c c c c c c }
 &  &  & 1 & 0 & \ldots &  &  & 1 & 0 & \ldots &  &  & 1 & 0 & \ldots &  &  & 1 & 0 & 1 & \ldots &  &  & 1 & 0 & 1 & \ldots &  &  & 1 & 0 & 1 \\
+ &  &  &  &  &  &  &  &  &  &  &  &  &  &  &  &  &  &  &  &  &  &  &  &  &  &  &  &  &  &  &  &  & \\
 &  &  & 1 & 1 & \ldots &  &  & 1 & 1 & \ldots &  &  & 1 & 1 & \ldots &  &  & 1 & 1 & 1 & \ldots &  &  & 1 & 1 & 1 & \ldots &  &  & 1 & 1 & 1 \\
+ &  &  &  &  &  &  &  &  &  &  &  &  &  &  &  &  &  &  &  &  &  &  &  &  &  &  &  &  &  &  &  &  & \\
 &  &  & 1 & 1 & \ldots &  &  & 1 & 1 & \ldots &  &  & 1 & 1 & \ldots &  & 1 & 0 & 0 & 0 & \dots &  & 1 & 0 & 0 & 0 & \ldots &  & 1 & 0 & 0 & 0 \\
\hline
& 1 & 0 & 0 & 0 & \ldots & 1 & 0 & 0 & 0 & \ldots & 1 & 0 & 0 & 0 & \ldots & 1 & 0 & 1 & 0 & 0 & \ldots & 1 & 0 & 1 & 0 & 0 & \ldots & 1 & 0 & 1 & 0 & 0.
\end{tabular}}
\end{center}

If we subtract the sum of the sizes of the elements in $A'$ from the last sum, we get the sum o the sizes of the elements in  $A \setminus A'$:
\begin{center}
\resizebox{\textwidth}{!}{\begin{tabular}{ c c c c c c c c c c c c c c c c c c c c c c c c c c c c c c c c c c }
& 1 & 0 & 0 & 0 & \ldots & 1 & 0 & 0 & 0 & \ldots & 1 & 0 & 0 & 0 & \ldots & 1 & 0 & 1 & 0 & 0 & \ldots & 1 & 0 & 1 & 0 & 0 & \ldots & 1 & 0 & 1 & 0 & 0\\
- &  &  &  &  &  &  &  &  &  &  &  &  &  &  &  &  &  &  &  &  &  &  &  &  &  &  &  &  &  &  &  &  & \\
&  & 1 & 0 & 0 & \ldots &  & 1 & 0 & 0 & \ldots &  & 1 & 0 & 0 & \ldots &  & 1 & 0 & 1 & 0 & \ldots &  & 1 & 0 & 1 & 0 & \ldots &  & 1 & 0 & 1 & 0 \\
\hline
&  & 1 & 0 & 0 & \ldots &  & 1 & 0 & 0 & \ldots &  & 1 & 0 & 0 & \ldots &  & 1 & 0 & 1 & 0 & \ldots &  & 1 & 0 & 1 & 0 & \ldots &  & 1 & 0 & 1 & 0,
\end{tabular}}
\end{center}
so we conclude that the PARTITION instance output by the reduction is indeed a positive instance.

Conversely, assume there exists a partition $(A',A \setminus A')$ of $A$ such that the sums of the corresponding sizes are the same. That means
\[
\sum_{a \in A'} s(a) = \sum_{a \in A\setminus A'} s(a),
\]
and from the discussion above, each of these sums must be
\begin{center}
\resizebox{\textwidth}{!}{\begin{tabular}{ c c c c c c c c c c c c c c c c c c c c c c c c c c c }
1 & 0 & 0 & \ldots & 1 & 0 & 0 & \ldots & 1 & 0 & 0 & \ldots & 1 & 0 & 1 & 0 & \ldots & 1 & 0 & 1 & 0 & \ldots & 1 & 0 & 1 & 0.
\end{tabular}}
\end{center}

If it were the case that both $s(b_{1})$ and $s(b_{2})$ belonged to $A'$, then
\begin{align*}
\sum_{a \in A'} s(a) &\geq s(b_{1}) + s(b_{2})\\
& > \sum_{a \in A\setminus \{b_{1}, b_{2}\}} s(a)\\
& \geq \sum_{a \in A\setminus A'} s(a),
\end{align*}
because $s(b_{1})  + s(b_{2})$ is given by
\begin{center}
\resizebox{\textwidth}{!}{\begin{tabular}{ c c c c c c c c c c c c c c c c c c c c c c c c c c c c c }
 &  & 1 & 1 & \ldots &  & 1 & 1 & \ldots &  & 1 & 1 & \ldots &  & 1 & 1 & 1 & \ldots &  & 1 & 1 & 1 & \ldots &  & 1 & 1 & 1 \\
+ &  &  &  &  &  &  &  &  &  &  &  &  &  &  &  &  &  &  &  &  &  &  &  &  &  &  & \\
 &  & 1 & 1 & \ldots &  & 1 & 1 & \ldots &  & 1 & 1 & \ldots & 1 & 0 & 0 & 0 & \dots & 1 & 0 & 0 & 0 & \ldots & 1 & 0 & 0 & 0 \\
\hline
& 1 & 1 & 0 & \ldots & 1 & 1 & 0 & \ldots & 1 & 1 & 0 & \ldots & 1 & 1 & 1 & 1 & \ldots & 1 & 1 & 1 & 1 & \ldots & 1 & 1 & 1 & 1,
\end{tabular}}
\end{center}
while $\sum_{a \in A\setminus \{b_{1}, b_{2}\}} s(a)$ is
\begin{center}
\resizebox{\textwidth}{!}{\begin{tabular}{ c c c c c c c c c c c c c c c c c c c c c c c c c c c c c c c c c c }
& 1 & 0 & 0 & 0 & \ldots & 1 & 0 & 0 & 0 & \ldots & 1 & 0 & 0 & 0 & \ldots & 1 & 0 & 1 & 0 & 0 & \ldots & 1 & 0 & 1 & 0 & 0 & \ldots & 1 & 0 & 1 & 0 & 0\\
- &  &  &  &  &  &  &  &  &  &  &  &  &  &  &  &  &  &  &  &  &  &  &  &  &  &  &  &  &  &  &  &  & \\
&  & 1 & 1 & 0 & \ldots &  & 1 & 1 & 0 & \ldots &  & 1 & 1 & 0 & \ldots &  & 1 & 1 & 1 & 1 & \ldots &  & 1 & 1 & 1 & 1 & \ldots &  & 1 & 1 & 1 & 1\\
\hline
&  &  & 1 & 0 & \ldots &  &  & 1 & 0 & \ldots &  &  & 1 & 0 & \ldots &  &  & 1 & 0 & 1 & \ldots &  &  & 1 & 0 & 1 & \ldots &  &  & 1 & 0 & 1.
\end{tabular}}
\end{center}
So, if both $s(b_{1})$ and $s(b_{2})$ belonged to $A'$, we would have a contradiction. It follows that $s(b_{1})$ and $s(b_{2})$ must belong to different parts of the partition. Assume without loss of generality that $b_{1} \in A'$ and $b_{2} \in A\setminus A'$.

Note that $\sum_{a \in A'}s(a) - s(b_{1})$ is
\begin{center}
\resizebox{\textwidth}{!}{\begin{tabular}{ c c c c c c c c c c c c c c c c c c c c c c c c c c c c }
& 1 & 0 & 0 & \ldots & 1 & 0 & 0 & \ldots & 1 & 0 & 0 & \ldots & 1 & 0 & 1 & 0 & \ldots & 1 & 0 & 1 & 0 & \ldots & 1 & 0 & 1 & 0\\
- &  &  &  &  &  &  &  &  &  &  &  &  &  &  &  &  &  &  &  &  &  &  &  &  &  & \\
&  & 1 & 1 & \ldots &  & 1 & 1 & \ldots &  & 1 & 1 & \ldots &  & 1 & 1 & 1 & \ldots &  & 1 & 1 & 1 & \ldots &  & 1 & 1 & 1 \\
\hline
&  &  & 1 & \ldots &  &  & 1 & \ldots &  &  & 1 & \ldots &  &  & 1 & 1 & \ldots &  &  & 1 & 1 & \ldots &  &  & 1 & 1.
\end{tabular}}
\end{center}

Now, define the set
\[
B' =  F^{-1}\left(D \cap \left(A'\setminus \left\{s(b_{1})\right\}\right)\right).
\]
This is the set of elements with  nonzero sizes in the input SUBSET-SUM instance that are mapped to the elements with nonzero sizes in $A'\setminus \left\{s\left(b_{1}\right\}\right)$ through $F$. The sum of the sizes of the elements in $B'$ is then
\begin{center}
\begin{tabular}{ c c c c c c c c c c c c c c }
1 & \ldots & 1 & \ldots & 1 & \ldots & 1 & 1 & \ldots & 1 & 1 & \ldots & 1 & 1,
\end{tabular}
\end{center}
 because the sum of the entries in each column is preserved. In conclusion, $\sum_{a\in B'} s(a) = t,$ so the SUBSET-SUM instance is a postitive instance. This finishes the argument showing $\text{SUBSET-SUM} \leq_{fop} \text{PARTITION}$.

\printbibliography 

@article{v013a004,
 author = {Murray, Cody D. and Williams, R. Ryan},
 title = {On the (Non) NP-Hardness of Computing Circuit Complexity},
 year = {2017},
 pages = {1--22},
 doi = {10.4086/toc.2017.v013a004},
 publisher = {Theory of Computing},
 journal = {Theory of Computing},
 volume = {13},
 number = {4},
 URL = {https://theoryofcomputing.org/articles/v013a004},
}

@book{10.5555/578533,
author = {Garey, Michael R. and Johnson, David S.},
title = {Computers and Intractability: A Guide to the Theory of NP-Completeness},
year = {1979},
isbn = {0716710447},
publisher = {W. H. Freeman \& Co.},
address = {USA}
}

@article{10.1007/BF01195200,
author = {Stewart, Iain A.},
title = {On completeness for NP via projection translations},
year = {1994},
issue_date = {March/April 1994},
publisher = {Springer-Verlag},
address = {Berlin, Heidelberg},
volume = {27},
number = {2},
issn = {0025-5661},
url = {https://doi.org/10.1007/BF01195200},
doi = {10.1007/BF01195200},
journal = {Math. Syst. Theory},
month = mar,
pages = {125–157},
numpages = {33}
}

@book{SipserToC3rd,
author = {Sipser, Michael},
title = {Introduction to the Theory of Computation},
year = {2013},
publisher = {Cengage}
}

@book{ImmermanDC,
author = {Immerman, Neil},
title = {Descriptive Complexity},
year = {1999},
publisher = {Springer},
doi = {10.1007/978-1-4612-0539-5}
}

@article{doi:10.1137/0213028,
author = {Chandra, Ashok K. and Stockmeyer, Larry and Vishkin, Uzi},
title = {Constant Depth Reducibility},
journal = {SIAM Journal on Computing},
volume = {13},
number = {2},
pages = {423-439},
year = {1984},
doi = {10.1137/0213028},
URL = {https://doi.org/10.1137/0213028},
eprint = {https://doi.org/10.1137/0213028}
}

@article{doi:10.1137/S0097539794270236,
author = {Allender, Eric and Balc\'{a}zar, Jos\'{e} and Immerman, Neil},
title = {A First-Order Isomorphism Theorem},
journal = {SIAM Journal on Computing},
volume = {26},
number = {2},
pages = {557-567},
year = {1997},
doi = {10.1137/S0097539794270236},
URL = {https://doi.org/10.1137/S0097539794270236},
eprint = {https://doi.org/10.1137/S0097539794270236}
}

@MISC {170906,
    TITLE = {Is it known whether PARTITION is NP-complete via first order reductions?},
    AUTHOR = {Pa\'ul, Risco},
    HOWPUBLISHED = {Computer Science Stack Exchange},
    NOTE = {version: 2025-01-12},
    URL = {https://cs.stackexchange.com/q/170906}
}

\end{document}